\documentclass[]{article}

\RequirePackage[T1]{fontenc}

\usepackage[switch,columnwise]{lineno}

\RequirePackage{graphicx}
\RequirePackage{flushend}
\RequirePackage[numbers,sort&compress]{natbib}
\RequirePackage[]{hyperref}

\usepackage{authblk}

\begin{document}

\title{Influence of incoherent scattering on stochastic deflection of high-energy negative particle beams in bent crystals}

\author[1]{I.V. Kirillin}
\author[1,2]{N.F. Shul'ga}
\author[3]{L. Bandiera}
\author[3,4]{V. Guidi}
\author[3,4]{A. Mazzolari}

\affil[1]{Akhiezer Institute for Theoretical Physics, National Science Center ``Kharkov Institute of Physics and Technology'', Akademicheskaya Str.,~1, 61108 Kharkov, Ukraine}
\affil[2]{V.N. Karazin Kharkov National University, Svobody Sq. 4, 61022 Kharkov, Ukraine}
\affil[3]{INFN Sezione di Ferrara, Via Saragat 1 44122 Ferrara, Italy}
\affil[4]{Dipartimento di Fisica e Scienze della Terra, Universit{\`a} degli Studi di Ferrara, Via Saragat 1 44122 Ferrara, Italy}

\date{June 14, 2016}

\maketitle

\begin{abstract}
An investigation on stochastic deflection of high-energy negatively charged particles in a bent crystal was carried out.
On the basis of analytical calculation and numerical simulation it was shown that it exists a maximum angle at which most of the beam is deflected.
The existence of a maximum, which is taken in the correspondence of the \textit{optimal radius of curvature}, is a novelty with respect to the case of positively charged particles, for which the deflection angle can be freely increased by increasing the crystal length.
This difference has to be ascribed to the stronger contribution of incoherent scattering affecting the dynamics of negative particles that move closer to atomic nuclei and electrons.
We therefore identified the ideal parameters for the exploitation of axial confinement for negatively charged particle beam manipulation in future high-energy accelerators, e.g., ILC or muon colliders.
\end{abstract}

\section{Introduction}

In contrast to an amorphous medium, there are selected directions in a crystal along which the atoms are arranged to form strings and planes.
If a high-energy charged particle moves in crystal at a small angle to one of these directions, its motion is governed by the field of continuous potentials of crystal atomic strings or planes \cite{Lindhard} that form potential wells in which particles trajectories can be trapped.
In the case of a bent crystal, these potential wells gives the possibility to deflect the direction of motion of a charged beam.
There are three main mechanisms of beam deflection by a bent crystal.
The first one is planar channeling, which was predicted in \cite{Tsyganov1, Tsyganov2} and experimentally observed in \cite{expplan}.
In this mechanism, charged particles are captured by the field of bent atomic planes and thereby deflected by an angle equal to the angle of crystal bend.
Channeling may occur as the particle incidence angle with the crystal axes (planes) is lower than the critical angle introduced by Lindhard, $\psi_c=\sqrt{{2U_{0,ax}}/{pv}}$ ($\theta_c=\sqrt{{2U_{0,pl}}/{pv}}$), $U_0$ being the continuous potential well depth, $p$ and $v$ the particle momentum and velocity, respectively \cite{Lindhard}.
Planar channeling is more effective for positively than for negatively charged particles.
Indeed, negative particles are attracted by atomic nuclei, the scatter on which causes their passage from under- to above-barrier states, i.e., the particles are easier dechanneled.
A method to diminish the influence of incoherent scattering with nuclei to the deflection efficiency could be the usage of the coherent interactions of above-barrier particles, i.e., not channeled, with bent crystals.
This is possible with the help of the volume reflection mechanism, that was proposed in \cite{Taratin} and experimentally observed in \cite{expvr1} and \cite{expvr2}.
Under volume reflection, particles that found in above-barrier states with respect to the planar potential wells are deflected to the direction opposite to the direction of crystal bend.
Volume reflection is efficient for either positively or negatively charged particles but the deflection angles are quite small, being of the order of $\theta_c$, and thus much smaller than for planar channeling in a bent crystal.
The third mechanism called stochastic deflection consists of the scattering of above-barrier particles in the field of bent crystal atomic strings.
This mechanism was proposed in \cite{1991} and experimentally observed in \cite{expsd1} and \cite{expsd2} for positively and negatively charged particles, respectively.
Stochastic deflection is efficient for both charge signs of scattered particles and gives the opportunity to deflect most of the beam at the crystal bending angle as for channeling, with a deflection efficiency close to 100\%.

As for planar channeling, the contribution of incoherent scattering plays a crucial role in the dynamics of negative particles moving in a crystal in the regime of stochastic deflection.
However, the role of incoherent scattering in stochastic deflection has never been deepened.
In this paper, we investigated the incoherent scattering contribution in the stochastic deflection of negatively charged particles by means of analytical calculation and Monte Carlo simulation.
We also analyzed the dependence of the deflection efficiency from the crystal radius of curvature, defining an \textit{optimal radius of curvature}.
A possibility for exploitation of stochastic deflection in electron beam collimation was also examined.

\section{Analytical model for stochastic deflection of negative particles}

In \cite{1991}, on the basis of a numerical simulation of high-energy charged particle motion in a bent crystal, it was shown that if the angle between the momentum of the particle and the direction of a crystal axis is smaller than the critical angle of axial channeling $\psi_{c}$, particle can be deflected because of scattering in the field of bent atomic strings.
This deflection was later called \textit{stochastic deflection} due to the fact that trajectories of particles participating in such deflection are similar to chaotic trajectories and are extremely sensitive to initial conditions.

After some years the conditions that correspond to stochastic deflection were found and in \cite{1995} it was shown that the following relationship must be satisfied:
\begin{equation}
\label{eq:1995}
	\frac{ l L }{ R^2 } \leq \psi_c,{}
	\label{eq:GSh}
\end{equation}
where $L$ and $R$ are the thickness and the radius of curvature of the crystal, respectively, $l$ is the mean length of the path that the particle crosses during scattering on one atomic string in the direction that is parallel to this atomic string.
Condition (\ref{eq:GSh}) was obtained without an account of incoherent scattering of the particle with individual atoms in the crystal, thereby works much better for positively charged particles than for negative ones, because, due to attraction of atomic nuclei, negatively charged particles move in crystal closer to atomic strings than positively charged particles.
Consequently, condition (\ref{eq:1995}) should be modified in the case of negative particles by taking into account incoherent scattering with atomic nuclei and electrons.

\begin{figure}
\includegraphics[width = \columnwidth, trim={5cm 10cm 1.2cm 5.5cm}, clip]{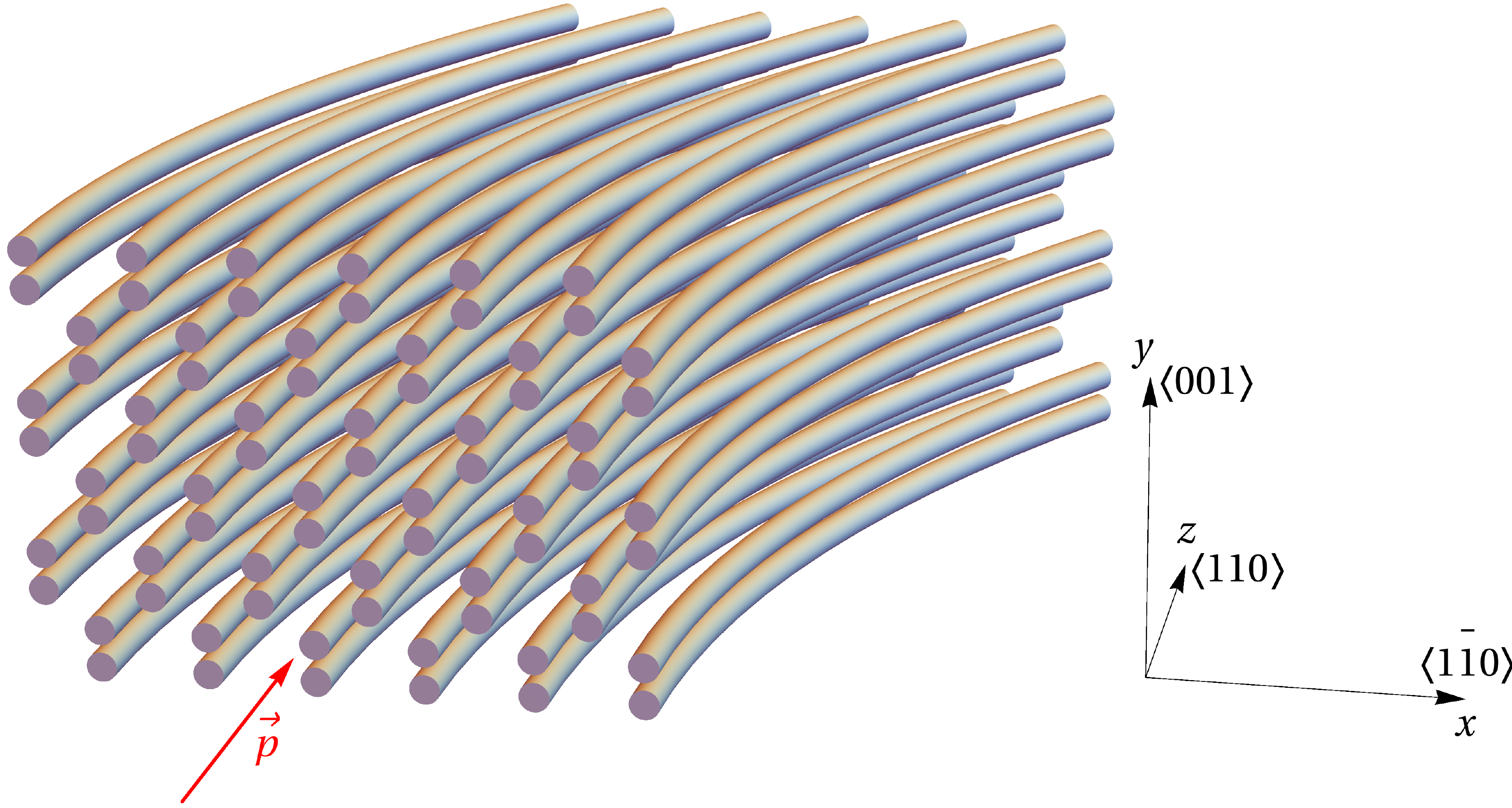}
\caption{The orientation of a Si crystal with respect to the impinging charged particles.}
\label{fig:axes}
\end{figure}

Let us consider the multiple scattering of above-barrier charged particle on atomic strings in a bent crystal step by step.
Each step corresponds to the scattering on one atomic string.
We denote crystal axis in the vicinity of which the particle impinges on the crystal as $z$-axis.
The initial angle, $\psi$, between the particle momentum and the $z$-axis, i.e. the incidence angle, is assumed to be less or a bit larger than the critical angle of axial channeling; we will define later in the text the right acceptance of stochastic deflection for negative particles with the support of computer simulations.
The $x$-axis is chosen to be orthogonal to the $z$-axis and lying in the $xz$ bending plane of the crystal (see Fig. \ref{fig:axes}).
The $y$-axis is orthogonal to the $x$-axis and $z$-axis.
During each scattering event on one atomic string, we consider the crystal as locally straight.
The curvature of the crystal and the incoherent scattering on thermal vibrations of atoms and electrons are taken into account when the particle passes from the scattering on one atomic string to scattering on a neighboring atomic string.
In such approximation, angular coordinates of the particle $\theta_x = v_x / v$ and $\theta_y = v_y / v$ (where $v_{x}$ and $v_{y}$ are projections of particle velocity $v$ on axes $x$ and $y$, respectively) after scattering on an atomic string can be written in the form of recurrent relations
\begin{eqnarray}
	\theta_{x,i+1} & = & \left( \theta_{x,i} - L_{i}/R \right) \cos\varphi_i - \theta_{y,i}\sin\varphi_i + \Psi_{x,i,inc} + L_i/R
	\label{eq:start1}\\
	\theta_{y,i+1} & = & \left( \theta_{x,i} - L_{i}/R \right) \sin\varphi_i + \theta_{y,i}\cos\varphi_i + \Psi_{y,i,inc},
	\label{eq:start2}
\end{eqnarray}
where $i$ is the sequence number of collision of the particle with an atomic string, $\varphi_{i}$ is the azimuthal scattering angle of the particle at $i$-th collision, $L_{i}$ is the length of the path the particle passed in the crystal before $i$-th collision, $\Psi_{x,i,inc}$ and $\Psi_{y,i,inc}$ are summands that correspond to incoherent scattering during $i$-th collision.
After performing the same procedure of averaging over $\varphi_{i}$ as in \cite{1995}, from (\ref{eq:start1}) and (\ref{eq:start2}) one can obtain:
\begin{equation}
	\frac{ d }{ dz }\overline{ \psi^2 } = l / R^2 + \frac{ d }{ dz }\overline{{ \Psi_{inc} }^2},
	\label{eq:diff}
\end{equation}
where $\overline{ \psi^2 }$ is the averaged value of the angle $\psi$ between particle momentum and the $z$-axis and $\overline{{ \Psi_{inc} }^2}$ is the mean square of the incoherent scattering angle.

If the angle between particle momentum and the axis of the crystal atomic string $\psi \ll 1$, then the mean length of the path that the particle passes during scattering on one atomic string $l$ can be obtained from the following relation \cite{Lindhard}:
\begin{equation}
	\frac {1} {l} = n d \psi \int_{-\infty}^{\infty} db \left( 1 - cos ( \varphi (b) ) \right),
	\label{eq:l}
\end{equation}
where $\varphi (b)$ is the scattering angle in the plane that is orthogonal to the axis of the string as a function of impact parameter $b$, $n$ is a concentration of atoms in the crystal, $d$ is the mean distance between neighboring atoms in the crystal atomic string.
The azimuthal deflection angle $\varphi (b)$ can be written in the form \cite{Akhiezer_fi}
\begin{equation}
	\varphi (b) = \pi - 2 b \int_{ \rho_0 }^{ \infty } {{d \rho} \over \rho^2} \left( 1 - {U_{st} (\rho) \over \epsilon_\perp} - {b^2 \over \rho^2} \right)^{-1/2},
	\label{eq:fi}
\end{equation}
$U_{st} (\rho)$ being the potential of the crystal atomic string and $\rho$ the distance from the string, $\rho_0$ being the minimal distance between the particle and the atomic string during scattering, $\epsilon_\perp$ is the energy of transverse motion, $E$ is the particle energy.
Since $\epsilon_\perp$ is a function of $\psi$, Eq. (\ref{eq:diff}) represents in general a non-linear differential equation.
However, in some special cases, Eq. (\ref{eq:diff}) can be integrated analytically.

If we choose the potential of a crystal atomic string in the form
\begin{equation}
U_{st} (\rho) = U_0 \left( \displaystyle \frac{a}{\rho} \right)^2,
\label{eq:U}
\end{equation}
where $U_0$ and $a$ are constants, then from (\ref{eq:l}) and (\ref{eq:fi}) one can obtain the following expression for $l$
\begin{equation}
	\frac {1} {l} = 4 n d \psi a I \sqrt {\frac {U_0} {\epsilon_\perp }},
	\label{eq:l1}
\end{equation}
where $\displaystyle I = \int_0^1 dt \cos^2 \left( \frac{\pi t}{2} \right) \left( 1 - t^2 \right)^{-3/2} \approx 0.7$.
Far from the string $\epsilon_\perp \approx \displaystyle \frac {E \psi^2}{2}$, therefore
\begin{equation}
	l \approx \frac {1} {4 n d a} \sqrt {\frac {E} {U_0}}.
	\label{eq:l2}
\end{equation}
Thereby for the potential (\ref{eq:U}) $l$ does not depend from the angle $\psi$ and hence we can integrate Eq. (\ref{eq:diff}) obtaining
\begin{equation}
	\overline{ \psi^2 } = l L / R^2 + \overline{{ \Psi_{inc} }^2}.
	\label{eq:end}
\end{equation}
Eq. (\ref{eq:end}) represents the averaged value of the angle $\psi$ between particle momentum and the $z$-axis.
The second term of the sum in Eq. (\ref{eq:end}), $\overline{{ \Psi_{inc} }^2}$, depends on the incoherent scattering angle and it is a novelty as compared to the case of positive particles \cite{1995}.
In analogy with the case of an amorphous medium \cite{PDG}, we assume that the mean square angle of incoherent multiple scattering is proportional to the thickness of the target: $\overline{{ \Psi_{inc} }^2}= \xi L$, where $\xi$ is a constant.
If we label the maximum value of $\psi$ for which particle take part in stochastic deflection as $\psi_{m}$, from Eq. (\ref{eq:end}) we obtain the espression for the crystal thickness, $L_{st}$, up to which negatively charged particles are deflected by stochastic scattering with bent crystal strings:
\begin{equation}
	L_{st} = \frac {\psi_{m}^{2}} {l / R^{2} + \xi}.
	\label{eq:L}
\end{equation}
From Eq. (\ref{eq:L}) follows that the stochastic deflection mechanism gives the possibility to deflect a beam of negative particles up to the maximum angle
\begin{equation}
	\alpha_{st} = \frac {L_{st}}{R} = \frac {\psi_{m}^{2}} {l / R + \xi R}.
	\label{eq:alp}
\end{equation}
If $\xi \approx 0$, Eq. (\ref{eq:alp}) would have the same form as in \cite{1995} where stochastic deflection of positive particles was introduced.
Indeed, due to the repulsive force of the nuclei, positively charged particles move quite far from atomic strings and therefore the incoherent contribution can be neglected.
In such a case, the larger the curvature radius the larger the maximum possible deflection angle, i.e. $\alpha_{st} \propto R$.
As a results, the crystal thickness $L_{st}$ up to which the beam of positively charged particles is deflected via stochastic mechanism is proportional to $R^{2}$.

For negatively charged particles $\xi \neq 0$ and $\alpha_{st}$ shows a \textit{maximum} for some radius of curvature.
To determine this optimal radius, $R_{opt}$, one need to equate to zero the derivative of $\alpha_{st}$ with respect to $R$, thus obtaining
\begin{equation}
	R_{opt} = \sqrt {l / \xi},
	\label{eq:ropt}
\end{equation}
which is valid in the approximation of atomic string potential in the form of Eq. (\ref{eq:U}).

By the analogy with an amorphous medium, $\xi$ should be proportional to $E^{-2}$.
This means that the optimal radius of curvature for the model (\ref{eq:U}) should linearly increase with the increase of the beam energy.

For more realistic models of the axial potential, Eq. (\ref{eq:diff}) can not be integrated analytically.
For this reason, with the aim of finding the correct value of the maximum angle $\alpha_{st}$ and of the optimal $R_{opt}$ we developed a Monte Carlo for the case of axial potential in Doyle-Turner approximation \cite{Doyle} and presented the results in the next section.

\section{Monte Carlo simulations and discussion}

In Ref. \cite{joint2015}, the concept of the relaxation length $l_{e}$ of the particles under axial confinement was introduced.
This term was used to determine the thickness of the crystal at which the number of particles that are stochastically deflected is $1/e$ times with respect to the number of particles that impinged on the crystal aligned to crystal strings.
In other words, for a crystal bent to a curvature radius $R$, the relaxation length $l_{e}$ determines the maximum crystal length for efficient steering of particles at the full bending angle.
In \cite{joint2015}, the dependence of $l_{e}$ vs. $R$ was found for 400 GeV/$c$ protons.
This dependence had the form close to a parabola which is consistent with the equation (\ref{eq:L}) in the case $\xi \approx 0$.

To analyze the same dependence of $l_{e}$ from $R$ for the case of negatively charged particles, we carried out numerical simulation of 150 GeV/$c$ $\pi^{-}$-mesons dynamics in the field bent atomic strings that are parallel to the $\langle 110 \rangle$ silicon crystal axis.
In the simulation the bending plane of the crystal coincided with the (001) crystal plane, while beam divergence was set equal to zero as in Fig. \ref{fig:axes}.
The orientation of the crystal with respect to the impinging charged particles is shown in Fig. \ref{fig:axes}.
The Monte Carlo code was the same as in Ref. \cite{plb,ncc,joint2015}.
The code solves the equation of motion in the field of continuum atomic string potential through numerical integration and also takes into account the contribution of incoherent scattering with atomic nuclei and electrons.

\begin{figure}
\includegraphics[width = \columnwidth]{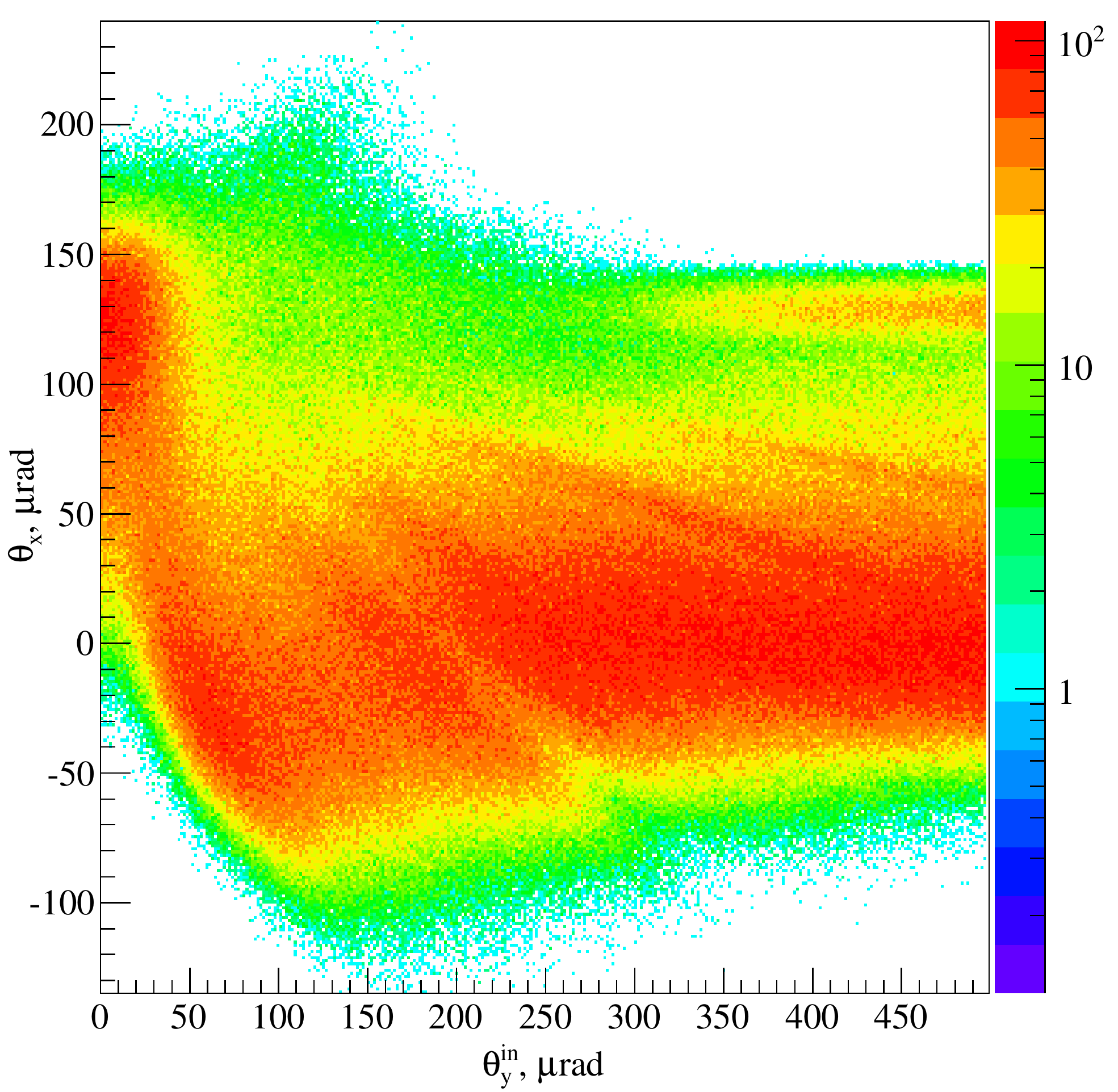}
\caption{The dependence of deflection angle of 150 GeV/$c$ $\pi^{-}$-mesons in the direction of crystal bend from the angle between particle initial momentum and the bending plane.}
\label{fig:scan}
\end{figure}

First of all we investigated the angular acceptance of the stochastic deflection mechanism for negative particles to find the value of maximum incidence angle, $\psi_{m}$.
Fig. \ref{fig:scan} displays the dependence of deflection angle of 150 GeV/$c$ $\pi^{-}$-mesons in the direction of crystal bend from the incidence angle $\theta_{y}^{in}$ between particle initial momentum and the bending plane (and thereby the $\langle 110 \rangle$ crystal axis which is contained in this plane).
The angle $\theta_{x}^{in}$ between particle initial momentum and the $(1\bar{1}0)$ plane was zero.
The length of the crystal was $L=1.52$ mm and the radius of curvature $R = 11.7$ m (this selection of crystal parameters will be explained hereinafter).
In the figure one can see that stochastic deflection mechanism efficiently works for negatively charged particles with $\theta_{y}^{in} \leq \psi_{c} \approx 37.4$ $\mu$rad and almost do not deflect particles with $\theta_{y}^{in} > 1.5 \psi_{c}$ to the direction of crystal bend.
For this reason, in the Monte Carlo we assumed that a particle is escaped from the stochastic deflection if the angle between its momentum and the current direction of atomic string exceeds $\psi_{m} = 1.5 \psi_{c}$.

Furthermore, in Fig. \ref{fig:scan} we can notice that when $\theta_{y}^{in}$ becomes higher than $\psi_{c}$, particles are deflected in the direction opposite to the direction of crystal bend.
This region corresponds to the doughnut scattering of particles in a bent crystal.
Moreover, when $\theta_{y}^{in} \gg \psi_{c}$ planar channeling occurs.
It is clearly visible in Fig. \ref{fig:scan} that the efficiency of planar channeling is much lower than the stochastic deflection one.
Indeed, because in the case of planar channeling particles can be deflected only if they are in under-barrier states, this mechanism of deflection is strongly spoiled by the incoherent scattering with nuclei that leads to the transition of particles from the under-barrier to above-barrier states.
On the contrary, under stochastic deflection above-barrier particles are deflected, explaining the less sensitivity to incoherent scattering.
\begin{figure}
\includegraphics[width = \columnwidth, trim={0.13cm 0.12cm 0.1cm 0.2cm}, clip]{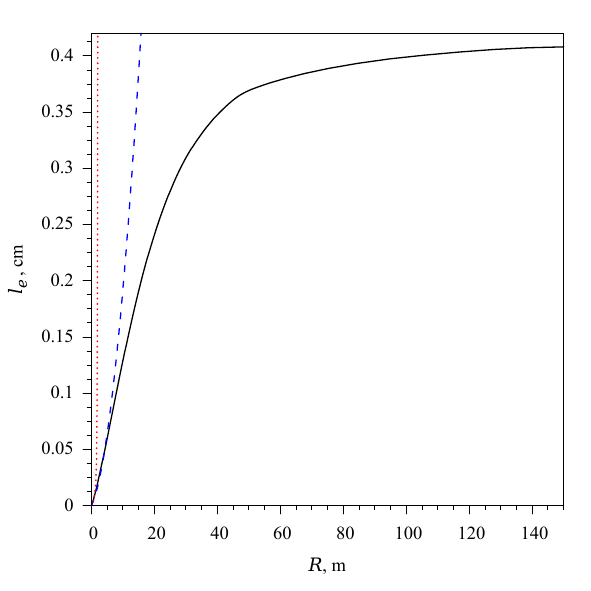}
\caption{Dependence of the relaxation length from the radius of curvature of the crystal for 150 GeV/$c$ $\pi^{-}$-mesons.
Ideal crystal (dotted line) with $\xi = 0$; Above-barrier particles motion with $\xi = 0$ (dashed line); Real crystal (solid line).}
\label{fig:l}
\end{figure}

Finally, the dependence of the relaxation length from the radius of curvature is displayed in Fig. \ref{fig:l}-solid curve, corresponding to particle motion in real crystal.
One can see that for small bending radii, the relaxation length in real crystal grows fast with $R$, while for larger radii the speed of growth of $l_{e}$ decreases tending to a constant value.
The constant value of $l_{e}$ corresponds to the case when $l / R^{2} \ll \xi$ in Eq. (\ref{eq:L}), therefore to the case in which $R \gg \sqrt{l / \xi}$.

\begin{figure}
\includegraphics[width = \columnwidth, trim={0.13cm 0.4cm 0.42cm 0.55cm}, clip]{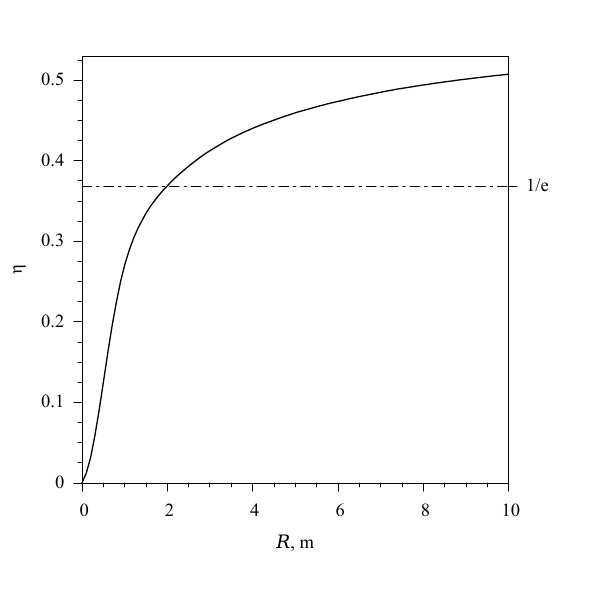}
\caption{Dependence of the fraction of channeled particles from the radius of curvature of the crystal.}
\label{fig:podbar}
\end{figure}

In real crystal we could not neglect the influence of the incoherent scattering to the particle dynamics in crystal.
Nevertheless, the model of ideal crystal could be used to analyze the dependence of $l_e$ in the case $\xi = 0$, represented in Fig. \ref{fig:l}-dotted curve.
We can notice that the relaxation length does not show the expected parabolic dependence from the radius of curvature as in Eq. (\ref{eq:L}).
Instead of this one can see that starting from $R \approx 2$ m $l_e$ grows with a vertical line dependence.
Indeed, in a straight crystal if a particle impinges on the crystal in the point where $U (x^{in},y^{in}) < 0$, it becomes under-barrier, i.e. axially channeled, and if we switched off the contribution of incoherent scattering, such particle remains in under-barrier state indefinitely.
In reality, since the axial dechanneling length of 150 GeV/$c$ $\pi^{-}$-meson is about 56 $\mu$m, particles that found initially in channeling mode are quickly dechanneled entering into the stochastic deflection regime.
In a bent crystal, because of curvature, the ratio of axially channeled particles is less than in a straight crystal, since the transverse energy $\epsilon_\perp$ contains the term $\propto E / R$.
The dependence of the ratio $\eta$ of under-barrier 150 GeV/$c$ $\pi^{-}$-mesons to the whole particles in the beam in the bend crystal aligned in accordance with Fig. \ref{fig:axes} is shown in Fig. \ref{fig:podbar}.
With the increase of $R$ this ratio increases and for $R = \infty$ (straight crystal) $\eta \approx 0.59$.
From Fig. \ref{fig:podbar} we see that for $R > 2$ m the ratio $\eta > 1/e$ (see dash-dotted line) and, because in ideal crystal we do not take into account incoherent scattering, starting from $R \approx 2$ m at least $1/e$ particles are deflected in axial channeling mode at the full bending angle of the crystal, i.e., $l_e$ goes to $\infty$ (see Fig. \ref{fig:l}-dotted curve).

As a consequence, we could not analyze the case $\xi = 0$ for stochastic deflection by considering the beam impinging on an ideal crystal because of the residual contribution of axial channeling.
However, if we analyze only above-barrier particles in an ideal crystal, all those particles are initially under the stochastic deflection regime and, according to Eq. (\ref{eq:L}), one expects a parabolic dependence of relaxation length from radius of curvature.
To prove this, in numerical simulation we selected for each particle with $U (x^{in},y^{in}) < 0$, $v_x^{in}$ and $v_y^{in}$ in such way that $\epsilon_\perp = 1$ eV.
On the other hand, for particles with $U (x^{in},y^{in}) > 0$, $v_x^{in}$ and $v_y^{in}$ were chosen equal to zero.
The result of simulation is displayed in Fig. \ref{fig:l}-dashed curve, which show a parabolic dependence of $l_e$ from $R$, well representing the case of motion of over-barrier positive particles under stochastic deflection regime \cite{joint2015}.

Summarizing, we investigated the dependence of the relaxation length on the radius of curvature in three notable cases.
Fig. \ref{fig:l}-dotted line represents the case of ideal crystal without contribution of incoherent scattering ($\xi = 0$) for which the contribution of axial channeling cannot be neglected; Fig. \ref{fig:l}-dashed line represents the case of above-barrier particles motion neglecting the contribution of incoherent scattering that well describes the case of positive particles motion in the field of bent atomic strings; Fig. \ref{fig:l}-solid line represents the case of real crystal in which the contribution of incoherent scattering limits the crystal length crossed by a negative particle under stochastic deflection regime.

\begin{figure}
\includegraphics[width = \columnwidth]{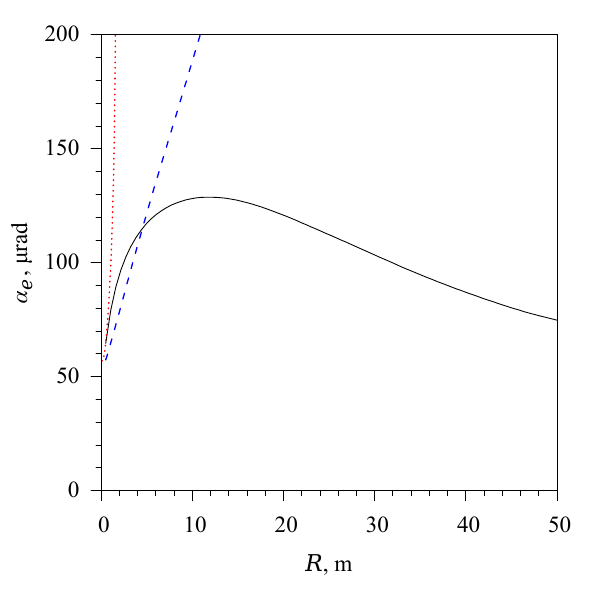}
\caption{Dependence of the deflection angle $\alpha_{e}$ from the radius of curvature $R$, for 150 GeV/$c$ $\pi^{-}$-mesons.
Ideal crystal (dotted line) with $\xi = 0$; Above-barrier particles motion with $\xi = 0$ (dashed line); Real crystal (solid line).}
\label{fig:alpha}
\end{figure}

\begin{figure}
\includegraphics[width = \columnwidth]{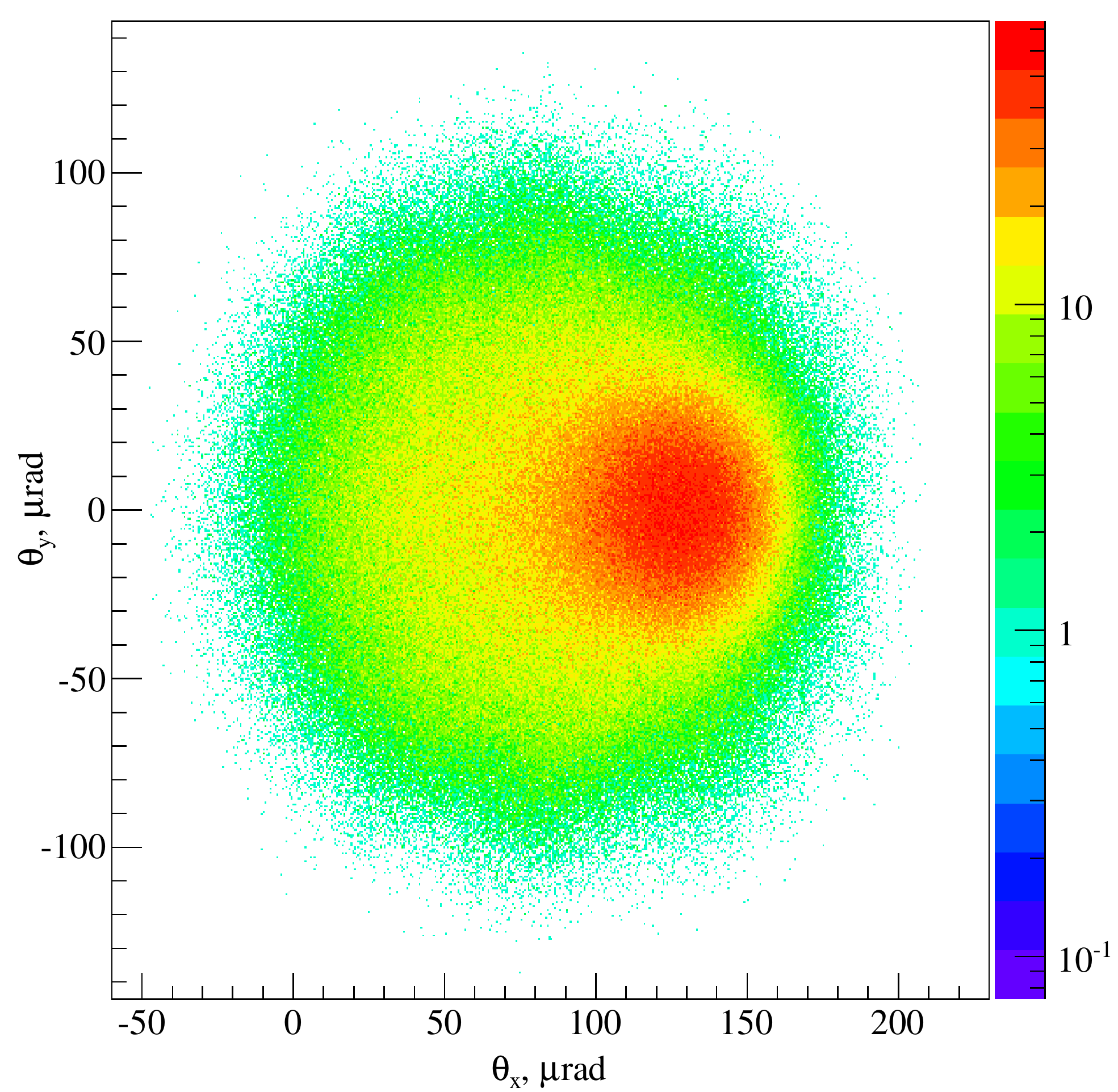}
\caption{Angular distribution of 150 GeV/$c$ $\pi^{-}$-mesons after passing bent Si crystal with $R = R_{opt}$ and $L = l_{e}$}
\label{fig:angdistr}
\end{figure}

In order to find the optimal radius of curvature, we analyzed via simulation the dependence of the deflection angle, at which $\frac{1}{e}$-th part of beam particles is deflected, over the radius of curvature.
The dependence of this deflection angle, $\alpha_{e} = l_{e} / R$, from the radius of curvature is plotted in Fig. \ref{fig:alpha}.
As in Fig. \ref{fig:l}, the black solid curve corresponds to particle beam motion in a real crystal, the red dotted curve correspond to beam motion in the ideal crystal and the blue dashed curve corresponds to above-barrier particle motion in the ideal crystal.
One can notice that the solid curve in Fig. \ref{fig:alpha} has a maximum at $R = R_{opt} \approx 11.7$ m (that is why this radius of curvature and $l_e$ that corresponds to it were chosen to obtain results shown on Fig. \ref{fig:scan}).
The existence of a maximum in the curve of Fig. \ref{fig:alpha} is explained by the plateau of the relaxation length, $l_e$, at large $R$ in Fig. \ref{fig:l}.
Indeed, this results in the impossibility of increasing the bending angle at large $R$ while simultaneously increasing the crystal length since $L$ becomes too larger than $l_e$.
This is the result of the strong contribution of incoherent scattering with nuclei and electrons on particle dynamics in crystals for the case of negative particles.
Conversely, for the case of positive particles (see \cite{joint2015}), $l_e$ grows with $R^2$ and hence $\alpha_{e}$ grows with $R$.
The same linear growth we see for the case of the above-barrier beam in the ideal crystal (dashed line in Fig. \ref{fig:alpha}).

Coherent interactions of charged particle beams and crystals have been strongly investigated for crystal-assisted positively charged beam manipulation.
In particular, it has been proposed to exploit planar channeling for the LHC proton beam collimation and extraction.
Since its low deflection efficiency (see Fig. \ref{fig:scan}), planar channeling has limited applications in the case of negative particles.
With the aim of improving the steering efficiency for negatively charged beams, stochastic deflection can be exploited.
As an example, Fig. \ref{fig:angdistr} displays the simulated 2D-angular distribution of 150 GeV/$c$ $\pi^{-}$-mesons after passage through a bent Si crystal with $R = R_{opt}$ and $L = l_{e}$ under stochastic deflection.
An efficient deflection of negatively charged particles at the angle of crystal bend, which is approximately equal to 130 $\mu$rad, is clearly visible.
Moreover, one can notice that almost all the beam particles are deflected to the direction of the curvature vector of the crystal with a horizontal deflection angle $>$ 0.
The large deflection efficiency and wide angular acceptance make stochastic deflection a good candidate for crystal-assisted negative beam manipulation in future muon or electron-positron colliders.
For instance, the collimation system of the International Linear Collider will be mainly made by spoiler-absorber pairs for the extraction of halo particles.
The interaction of the $e^{\pm}$ beam with the cm-long spoilers may result in wakefield perturbations.
The particles deflection by a mm-long bent crystal replacing one of the spoiler would increase halo-cleaning efficiency as compared to the case of an amorphous spoiler \cite{seryi,ban}.
Moreover, electrons interacting with the axial potential are subject to stronger radiative losses as compared to the amorphous case, thus further improving the discrimination of halo particles that are later swept away by outstream magnet.

\section{Conclusions}

In summary, an investigation on the mechanism of stochastic deflection of axially confined ultra-high energy negative particles in bent crystals was carried out through analytical estimations and Monte Carlo simulations.
We determined the ideal parameters for exploitation of stochastic deflection as a tool for negative high-energy beam manipulation in accelerators.
In particular, we found that, differently to the case of positively charged particles, the beam deflection angle can not be freely enhanced by increasing the crystal length.
In fact, the multiple scattering of negative particles in the field of atomic strings is strongly spoiled by the incoherent scattering with atomic nuclei and electrons.
As a results, there exists an optimal bending radius for which most of the beam is deflected at a maximal angle trough stochastic deflection, a possibility that can be exploited for beam manipulation in future muon or electron-positron colliders.

\section{acknowledgements}
We recognized partial support of the INFN-CHANEL experiment, the National Academy of Sciences of Ukraine (Project No CO-7-1) and the Ministry of Education and Science of Ukraine (Project No. 0115U000473).


\begin{thebibliography}{99}

\bibitem{Lindhard}
J.~Lindhard. Danske Vid. Selsk. Mat. Fys. Medd. \textbf{34}:14 (1965)

\bibitem{Tsyganov1}
E.N. Tsyganov. Preprint Fermilab TM-682 (1976)

\bibitem{Tsyganov2}
E.N. Tsyganov. Preprint Fermilab TM-684 (1976)

\bibitem{expplan}
A. F. Elishev, N. A. Filatova et al. Phys. Lett. B \textbf{88}, 387 (1979)

\bibitem{Taratin}
A.M. Taratin, S.A. Vorobiev. Phys. Lett. A \textbf{119}, 425 (1987)

\bibitem{expvr1}
Yu. M. Ivanov, A. A. Petrunin et al. Phys. Rev. Lett. \textbf{97}, 144801 (2006)

\bibitem{expvr2}
Yu. M. Ivanov, N. F. Bondar' et al. JETP Lett. \textbf{84}, 372 (2006)

\bibitem{1991}
A.A. Grinenko, N.F. Shul'ga, Pis'ma Zh. Eksp. Teor. Fiz. \textbf{54}, 520 (1991)

\bibitem{expsd1}
W. Scandale, A. Vomiero, et al. Phys. Rev. Lett. \textbf{101}, 164801 (2008)

\bibitem{expsd2}
W. Scandale, A. Vomiero, et al. Phys. Lett. B \textbf{680}, 301 (2009)

\bibitem{1995}
N.F. Shul'ga, A.A. Greenenko, Phys. Lett. B \textbf{353}, 373 (1995)

\bibitem{Akhiezer_fi}
A.I. Akhiezer, V.I. Truten', N.F. Shul'ga. Phys. Rep. \textbf{203}, 289 (1991)

\bibitem{PDG}
V.L. Highland, Nucl. Instrum. Methods \textbf{129}, 497 (1975); Nucl.
Instrum. Methods \textbf{161}, 171 (1979).

\bibitem{Doyle}
P.A. Doyle, P.S. Turner. Acta Cryst. A \textbf{24}, 390 (1968)

\bibitem{joint2015}
L. Bandiera, A. Mazzolari, E. Bagli, G. Germogli, V. Guidi, A. Sytov, I.V. Kirillin, N.F. Shul'ga,
A. Berra, D. Lietti, M. Prest, D. De Salvador, E. Vallazza. Eur. Phys. J. C \textbf{76}, 80 (2016)

\bibitem{plb}
N.F. Shul'ga, I.V. Kirillin, V.I. Truten'. Phys. Lett. B \textbf{702}, 100 (2011)

\bibitem{ncc}
N.F. Shul'ga, I.V. Kirillin, V.I. Truten'. Nuovo Cim. C \textbf{34}, 425 (2011)

\bibitem{seryi}
A. Seryi, Nucl. Instrum. Methods Phys. Res., Sect. A \textbf{623}, 23 (2010).

\bibitem{ban}
L. Bandiera, et al., Phys. Rev. Lett., \textbf{111}, 255502 (2013).


\end{thebibliography}
\end{document}